\documentclass[onecolumn,preprintnumbers,12pt,amsmath,amssymb,nofootinbib,draft]{revtex4}

\def	\be	{\begin{equation}}
\def	\ee	{\end{equation}}
\def	\bqt	{\begin{quote}}
\def	\eqt	{\end{quote}}
\def	\a	{\alpha}

\def	\mn 	{\mu \nu}
\def	\pl	{\partial}
\def	\del	{\nabla}

\begin{document}

\title{No Open or Flat Bouncing Cosmologies in Einstein Gravity}

\author{Maulik Parikh}

\affiliation{Department of Physics and Beyond: Center for Fundamental Concepts in Science \\
Arizona State University, Tempe, AZ 85287, USA} 

\begin{abstract}
\begin{center}
{\bf Abstract}
\end{center}
\noindent
We show that bouncing open or flat Friedmann-Robertson-Walker cosmologies are inconsistent with worldsheet string theory to first approximation. Specifically, the Virasoro constraint translates to the null energy condition in spacetime at leading order in the alpha-prime expansion. Thus one must go beyond minimally-coupled Einstein gravity in order to find bounce solutions.
\end{abstract}

\maketitle

\section{Introduction}
\label{introduction}

\thispagestyle{empty}
\noindent

To evade the Big Bang singularity, cosmologists have often envisioned the existence of a cosmological bounce, whereby the scale factor of the universe has a nonzero minimum, with a contracting phase preceding the present expansionary era. This idea, which dates back at least to the ``Tolman wormhole" \cite{tolman}, has seen many contemporary reincarnations (see, e.g., \cite{novello,battefeld} and references therein).

However, to realize this scenario within the context of a flat or open Friedmann-Robertson-Walker universe, Einstein's equations require that the null energy condition be violated. Consider an FRW universe in $d+1$ spacetime dimensions sourced by a fluid with energy density $\rho$ and pressure $p$, and with cosmological constant $\Lambda$. Einstein's equations are
\begin{eqnarray}
H^2 & = & \frac{16 \pi G_N}{d(d-1)} \rho - \frac{k}{a^2} + \frac{2 \Lambda}{d(d-1)} \label{frw1} \\
\dot{H} & = & - \frac{8 \pi G_N}{d-1} (\rho + p) + \frac{k}{a^2} \; . \label{frw2}
\end{eqnarray}
Since $\dot{H} = \frac{\ddot{a}}{a} - H^2$, and $H$ vanishes at the bounce, the left-hand side of (\ref{frw2}) is equal to $\frac{\ddot{a}}{a}$ at the moment of the bounce. Thus, for $k \neq 1$, positivity of $\ddot{a}$, which is a sufficient condition for a bounce, requires that $\rho + p < 0$, in violation of the NEC. This too-quick argument has a tiny loophole: the strict positivity of $\ddot{a}$ at $t = 0$ (the moment of the bounce) is a sufficient but not a necessary condition for a bounce to exist. For example, a scale factor that near the bounce was approximately quartic, rather than quadratic, would have even $\ddot{a}$ vanishing at $t = 0$. To rule out hypothetical NEC-satisfying bounce solutions of this type and close the loophole, consider a Taylor-expansion of the scale factor about $t = 0$:
\be
a(t) \approx a_* + c t^{2n} + ... \; .
\ee
Here $a_*$ is the scale factor at the bounce and $c$ is a strictly positive constant; the case of interest is $n > 1$. Substituting into (\ref{frw2}) gives
\be
\rho + p \approx \frac{d-1}{8 \pi G_N} \left ( \frac{k}{a_*^2} - 2n(2n-1) c t^{2n-2} + ... \right ) \; .
\ee
Evidently, even though $\rho + p$ can be zero at the bounce (for $k = 0$), $\rho + p$ must be strictly negative for arbitrarily small, nonzero $t$ in order for a bounce solution to exist. The naive conclusion therefore stands: to obtain a bouncing flat or open FRW cosmological solution in Einstein gravity, it is necessary that the NEC be violated \cite{molinaparis}.\footnote{Of course, for $k = +1$, bounce solutions that satisfy the NEC are easy to find, the simplest example of which is just global de Sitter space with scale factor $a(t) = \frac{1}{\sqrt{b \Lambda}} \cosh (\sqrt{b \Lambda} t)$ where $b = 2/(d^2-d)$. To rule out this case as well, one needs the strong energy condition; however, we will not consider closed-universe cosmologies here.}

Thus, in Einstein gravity, the possibility of flat or open bouncing cosmologies is intimately connected to the status of the NEC. Let us therefore briefly review the energy conditions. The energy conditions play a vitally important role in general relativity. They are the main local constraints determining which solutions of Einstein's equations are physical; without the energy conditions, every metric would formally constitute an exact solution (at least locally) to Einstein's equations. Moreover, the validity of various energy conditions is the primary assumption in the singularity theorems \cite{Penrose:1965} as well as in no-go theorems that prohibit the traversability of wormholes, the creation of laboratory universes \cite{Farhi:1987}, and the building of time machines \cite{Hawking:1992}. Perhaps most importantly, the energy conditions are invoked in black hole thermodynamics, where they are used in proving that the area of a black hole event horizon, like entropy, always increases \cite{Hawking:1976}. The most fundamental of the energy conditions is the null energy condition which, in its covariant formulation, states that at every point in spacetime,
\be
T_{\mn} v^\mu v^\nu \geq 0 \; ,	\label{TNEC}
\ee
for any light-like vector, $v^\mu$. The NEC is the weakest of the energy conditions in that a violation of the NEC automatically implies a violation of the weak, dominant, and strong energy conditions. 

Now, the importance of the NEC stems essentially from the null version of Raychaudhuri's equation which governs the focussing of light rays. Raychaudhuri's equation states that a null geodesic congruence with affine parameter $\lambda$, expansion parameter $\theta$, and shear $\sigma_{ab}$ satisfies
\be
\frac{d \theta}{d \lambda} = - \frac{1}{2} \theta^2 - \sigma_{ab} \sigma^{ab} - R_{\mn} v^\mu v^\nu \; ,
\ee
where we have ignored a vorticity term for simplicity. Thus if
\be
R_{\mn} v^\mu v^\nu \geq 0 \; , \label{RNec}
\ee
then every term on the right in Raychaudhuri's equation is negative, which is the key requirement in proving a number of gravitational theorems. The problem is that, at least within the framework of general relativity, there is no compelling reason why (\ref{RNec}), known as the null or Ricci convergence condition, should hold. However, if Einstein's equations are used, then
\be
R_{\mn} v^\mu v^\nu \geq 0  \Leftrightarrow T_{\mn} v^\mu v^\nu \geq 0 \; . \label{Rnec-Tnec}
\ee
That is, if the null energy condition, (\ref{TNEC}), holds for matter, then the Ricci convergence condition, (\ref{RNec}), holds for geometry (and vice versa). Attention thus shifts to the matter sector, described by quantum field theory. However, although (\ref{TNEC}) does hold for all familiar forms of matter, it does not appear to have a rigorous derivation, say in the classical limit of quantum field theory. Consequently, the validity of the NEC has been called into question \cite{Barcelo:2002,Rubakov:2014}. Indeed, there are known nontrivial NEC-violating effective theories of matter, such as conformal galileons and ghost condensates, which do not have ghost or gradient instabilities or any other obvious pathologies as far as QFT goes. Cosmologists have exploited these new theories of matter to obtain bouncing cosmological solutions to Einstein's equations \cite{quintombounce,galileonbounce,Buchbinder:2007ad,ghostbounce,superbounce}.

Thus it appears that there is neither an origin in general relativity of (\ref{RNec}), nor an origin in quantum field theory of (\ref{TNEC}). In fact, this is not altogether surprising. The distinction between matter and gravity is not fundamental: in Kaluza-Klein compactifications, as well as in changes of conformal frame, we see that what appears to be in the matter sector can equally be regarded as being geometry, thereby transforming what we mean by the energy-momentum tensor which in turn transforms the energy conditions \cite{jones}. A statement like (\ref{TNEC}) that applies only to matter is unlikely to be fundamental. Moreover, the overall sign of the matter action does not have any meaning in quantum field theory. To obtain conditions like (\ref{TNEC}) or (\ref{RNec}), we should therefore look to a theory of both matter and gravity, for which the overall sign of the matter action does have meaning since it is now a relative sign, relative to that of the gravitational action.

Indeed, the Ricci convergence condition, (\ref{RNec}), has recently been derived as a generic consequence of string theory \cite{NECderivation}, a theory which of course does describe both matter and gravity. The purpose of this paper is to point out the implications of that result for cosmological bounces. An immediate corollary of the derivation is that open or flat bouncing cosmological solutions of Einstein's equations are not consistent with string theory. But the regimes in which the derivation is obtained also reveal how the no-go theorem might potentially be evaded. We will therefore briefly review the derivation.

\section{Derivation of the Ricci Convergence Condition}

Worldsheet string theory is described by a two-dimensional nonlinear sigma model in which $D$ scalar fields, $X^\mu (\sigma, \tau)$ -- we focus here on bosonic string theory -- are minimally coupled to two-dimensional Einstein gravity on the worldsheet. The Polyakov action for a string propagating in flat space is
\be
S_P[X^\mu,h_{ab}] = - \frac{1}{4 \pi \a'} \int d^2 \sigma \sqrt{-h} h^{ab} \pl_a X^\mu \pl_b X^\nu \eta_{\mn} - \frac{c}{4 \pi} \int d^2 \sigma \sqrt{-h}  R_h \; ,  \label{flat-action}
\ee
where $R_h$ is the worldsheet Ricci scalar, and $c$ is an arbitrary constant. The equation of motion for the worldsheet metric, $h_{ab}$, is just Einstein's equation in two dimensions:
\be
0 = T^{\rm ws}_{ab} \; . \label{wsE}
\ee
Here the left-hand side is zero because the Einstein tensor vanishes identically in two dimensions, and $T^{\rm ws}_{ab}$ is the energy-momentum tensor on the worldsheet:
\be
T^{\rm ws}_{ab} \equiv -\frac {2} {\sqrt{-h}} \frac{\delta S_P}{\delta h^{ab}} =  \frac{1}{2 \pi \a'} \left ( \pl_a X^\mu \pl_b X^\nu \eta_{\mn} - \frac{1}{2} h_{ab} (\pl X)^2 \right ) \; .
\ee
Switching to light-cone coordinates on the worldsheet,
\be
\sigma^{\pm} \equiv \tau \pm \sigma \; ,
\ee
we see that the diagonal components of the worldsheet metric vanish:
\be
h_{ab} = \left ( \begin{array}{lc} \; \; 0 \; \;  \; -\frac{1}{2} \\ -\frac{1}{2} \; \; \;  \; \; \; 0 \end{array} \right ) \; .
\ee
The two-dimensional Einstein equation (\ref{wsE}) then reads
\be
0 =  \pl_+ X^\mu \pl_+ X^\nu \eta_{\mn} \; ,
\ee
with a similar equation holding when $+$ is replaced by $-$. These are the Virasoro constraints. Defining a vector field $v_+^\mu = \pl_+ X^\mu(\sigma, \tau)$, we find that
\be
\eta_{\mn} v_+^\mu v_+^\nu = 0 \; , \label{nullv}
\ee
which is to say that $v_+^\mu$ is a null vector field. The important point here is that worldsheet string theory naturally singles out spacetime null vectors. 

Next we will see how these lead to the Ricci convergence condition when the string propagates in an arbitrary curved spacetime. The Polyakov action is now 
\be
S_P[X^\mu,h_{ab}] = - \frac{1}{4 \pi \a'} \int d^2 \sigma \sqrt{-h} h^{ab} \pl_a X^\mu \pl_b X^\nu g_{\mn}(X) - \frac{1}{4 \pi} \int d^2 \sigma \sqrt{-h} \Phi(X) R_h  \; . \label{curved-action}
\ee
We have replaced the Minkowski metric $\eta_{\mn}$ by the general spacetime metric $g_{\mn}(X)$. Consistent with worldsheet diffeomorphism-invariance, we have also allowed there to be a scalar field, $\Phi(X(\tau, \sigma))$, which is the dilaton field; we neglect the anti-symmetric Kalb-Ramond field, $B_{\mn}$, for simplicity.

We now perform a background field expansion $X^\mu(\tau, \sigma) = X^\mu_0 (\tau, \sigma) + Y^\mu (\tau, \sigma)$ where $X^\mu_0 (\tau, \sigma)$ is some solution of the classical equation of motion. Then, for every value of $(\tau, \sigma)$, we can use standard field redefinitions \cite{CallanThorlacius,GSW1} to expand the metric in Riemann normal coordinates about the spacetime point $X^\mu_0 (\tau, \sigma)$:
\be
g_{\mn} (X) = \eta_{\mn} - \frac{1}{3} R_{\mu \a \nu \beta} (X_0) Y^\a Y^\beta - \frac{1}{6} \del_\rho R_{\mu \a \nu \beta} (X_0) Y^\rho Y^\a Y^\beta + ... \; . \label{RNC}
\ee
Contracted with $\partial_a X^\mu \partial^a X^\nu$, the second and higher terms introduce quartic and higher terms in the Lagrangian; spacetime curvature turns (\ref{curved-action}) into an interacting theory. The resultant divergences can be cancelled by adding suitable counter-terms to the original Lagrangian. Integrating out $Y$, the one-loop effective action is
\cite{CallanThorlacius,GSW1}
\be
S[X_0^\mu,h_{ab}] = - \frac{1}{4 \pi \a'} \int d^2 \sigma \sqrt{-h} \left ( h^{ab} \pl_a X_0^\mu \pl_b X_0^\nu (\eta_{\mn} + C_\epsilon \a' R_{\mn} (X_0)) -  \a' C_\epsilon \Phi(X_0) R_h \right ) \; . \label{eff-action}
\ee
Here $C_\epsilon$ is the divergent coefficient of the counter-terms. 

In light-cone coordinates, the Virasoro constraint now reads
\be
0 =  \pl_{\pm} X_0^\mu  \pl_{\pm} X_0^\nu \left ( \eta_{\mu \nu} + C_\epsilon \alpha' R_{\mu \nu} + 2 C_\epsilon\alpha' \nabla_\mu \nabla_\nu \Phi \right ) \; . 
\ee
Consider then an arbitrary null vector $v^\mu$ in the tangent plane of some arbitrary point in an arbitrary spacetime. Let there be a test string passing through the given point with $\pl_+ X^\mu$ equal to $v^\mu$ at the point. Note that at every point $X^\mu_0 (\tau, \sigma)$ the spacetime metric $g_{\mn}(X)$ is just $\eta_{\mn}$; for a string passing through such a point, $\pl_+X^\mu$ is therefore null with respect to both $\eta_{\mn}$ and $g_{\mn}(X)$. 
Defining $v_+^\mu = \pl_+ X^\mu$ as before, we find (\ref{nullv}) at leading order in $\alpha'$. At the next order, we have
\be
v_+^\mu v_+^\nu (R_{\mn} + 2 \nabla_\mu \nabla_\nu \Phi  ) = 0 \; . \label{alphavirasoro}
\ee
This is tantalizingly close to our form of the Ricci convergence condition, (\ref{RNec}), except for two differences: it is an equality, rather than an inequality, and there is an additional, unwanted term involving the dilaton.

However, now we recall that the metric that appears in the worldsheet action is the string-frame metric. We can transform to Einstein frame by defining:
\be
g_{\mn} = e^{\frac{4 \Phi}{D-2}} g_{\mn}^E \; .
\ee
Then we find that
\be
R^E_{\mn} v_+^\mu v_+^\nu  = + \frac{4}{D-2} (v_+^\mu \nabla^E_\mu \Phi )^2 \; .
\ee
The right-hand side is manifestly non-negative. Hence we have
\be
R^E_{\mn} v_+^\mu v_+^\nu  \geq 0 \; .
\ee
This establishes the Ricci convergence condition, which is equivalent to the null energy condition when Einstein's equations hold.

In summary, the spacetime interpretation of the Virasoro constraint on the worldsheet is precisely the Ricci convergence condition \cite{NECderivation}. This result was obtained in the critical number of spacetime dimensions ($D = 26$ for a bosonic string) because the worldsheet one-loop effective action we used was for a string moving in the critical number of dimensions. But it is easy to show that the Ricci convergence condition also holds in lower dimensions. This is because, for a given compactification, it is straightforward to compute the lower-dimensional Ricci tensor in terms of the higher-dimensional Ricci tensor. When this is done, the upshot is that, for both Kaluza-Klein and warped compactifications, if the Ricci convergence condition holds in $D$ dimensions, then it holds in $D-1$ dimensions \cite{NECderivation}. Since we have already established the Ricci convergence condition in the maximal (critical) number of dimensions, it follows by induction that it holds in all lower dimensions.

\section{Implications for Bouncing Cosmologies}

We can now apply $R_{\mn} v^\mu v^\nu \geq 0$ directly to a $k \neq 1$ FRW cosmology. In standard Robertson-Walker coordinates, we have
\be
(v^t)^2 = a^2(t) \vec{v}^2 \; ,
\ee
where $v^a = (v^t, \vec{v})$ is any null vector. Then we find that
\be
R_{\mn} v^\mu v^\nu =  \vec{v}^2 a^2(t) (d-1) \left ( - \dot{H} + \frac{k}{a^2} \right ) \; .
\ee
Thus the Ricci convergence condition required by worldsheet string theory establishes that, for $k \neq 1$, $\dot{H}$ cannot be positive, thereby precluding a bounce solution. In summary, consistency with string theory rules out open or flat bouncing FRW solutions in Einstein gravity, independently of (so far unprovable) assumptions about matter satisfying the null energy condition. The derivation of the Ricci convergence condition in string theory (or in its low-energy limit, supergravity) shuts the door on the simplest kinds of bounce cosmologies, in which some exotic NEC-violating matter field is minimally coupled to Einstein gravity. Such solutions are not consistent with low-energy string theory.

But the derivation also reveals the limitations of the claim. In particular, we derived the Ricci convergence condition while working to the lowest order in alpha prime and to the lowest genus in the string genus expansion. In field theory language, these two regimes correspond to working at leading order in the curvature expansion and at tree-level in the loop expansion. Consider first the effect of radiative corrections. It is tempting to write down the semi-classical Einstein equations:
\be
G_{\mn} = 8 \pi G_N \langle T_{\mn} \rangle - \Lambda g_{\mn} \; , \label{semiclassical}
\ee
where $\langle T_{\mn} \rangle$ is the renormalized stress tensor of matter calculated in some state in a given background. There is, to our knowledge, no rigorous derivation of such an equation starting from a theory that treats both gravity and matter quantum-mechanically. However, if such an equation were to exist, it would offer the possibility of bounce solutions. The left-hand side would still lead to the left-hand sides of (\ref{frw1}) and (\ref{frw2}), but the right-hand side would presumably no longer obey the null energy condition even for matter that obeys it classically. Indeed, the Casimir energy even for ordinary fields can cause $\langle T_{\mn} \rangle$ to violate the NEC. Although (\ref{semiclassical}) appears plausibly true as a semi-classical approximation, it is not obvious that it is consistent with string theory either. This is because (\ref{semiclassical}) treats gravity classically while treating matter quantum-mechanically. Yet in closed string theory, the graviton is just another mode of the string, so such an unequal treatment of gravity and matter is in tension, at the minimum, with the spirit of string theory.

The other, perhaps more promising, direction is to modify the geometric part of Einstein's equations. There are several ways this can be used to find bounces, such as non-minimal coupling to matter \cite{prebigbang} including kinetic-gravity braiding \cite{G-bounce}, massive gravity \cite{langlois}, or the inclusion of torsion \cite{torsionbounce}. The most obvious extension is to add higher-curvature terms to the gravitational equations \cite{stringinspiredbounces,Brandenberger:1993ef}. The presence of these terms is expected not only because Einstein gravity is not renormalizable, but also because the low-energy effective action of string theory naturally contains such terms at higher order in the alpha-prime expansion. When higher-curvature terms are included, two things happen. First, the left-hand side of (\ref{alphavirasoro}) contains additional terms. Thus, the Virasoro constraint on the worldsheet no longer translates to the Ricci convergence condition in spacetime, which is a necessary pre-condition if flat or open FRW cosmologies are to have bounces. Second, since the geometric part of Einstein's equations is modified, (\ref{Rnec-Tnec}) no longer holds; any new terms in the gravitational equations that are not proportional to the metric break the equivalence between (\ref{RNec}) and (\ref{TNEC}). NEC violation then becomes neither necessary nor sufficient to having a cosmological bounce solution to the gravitational equations of motion. Consequently, a bounce could potentially be achieved by coupling gravity to ordinary matter, or even perhaps to no matter at all. This is helpful because at present the known NEC-violating theories of matter are themselves only non-renormalizable effective theories; they do not have a clear origin within an ultraviolet-complete theory.

Finally, there could be implicit assumptions in our proof. (Of course one such assumption is that the correct theory of matter and gravity is string theory.) For example, in (\ref{RNC}) we assumed that the metric can be expanded about the spacetime point $X^\mu_0(\tau, \sigma)$ in Riemann normal coordinates, which is the case if there is an open coordinate patch around the point. However, as noted in \cite{NECderivation}, that is not true if the point lies on a boundary of spacetime. Thus potential NEC violations may be allowed if spacetime has a boundary. It is noteworthy in this context that string theory contains orientifolds, extended objects with negative tension which violate the NEC and which are non-dynamical branes that typically live on a boundary of spacetime. These may have interesting applications to cosmology \cite{savmartinec}, including the search for viable bounce solutions \cite{miguel1,miguel2}.

\bigskip
\noindent
{\bf Acknowledgments}

\noindent
M. P. is supported in part by DOE grant DE-FG02-09ER41624.


\begin{thebibliography}{99}

\bibitem{tolman}
R.~C.~Tolman, ``On the Theoretical Requirements for a Periodic Behaviour of the Universe," Phys. Rev. {\bf 38}, 1758 (1931).

\bibitem{novello}
M.~Novello and S.~E.~Perz Bergliaffa, ``Bouncing Cosmologies," Phys. Rept. {\bf 463}, 127 (2008); {\tt arXiv:0802.1634 [astro-ph]}.
  
\bibitem{battefeld}
D.~Battefeld and P.~Peter, ``A Critical Review of Classical Bouncing Cosmologies," {\tt arXiv:1406.2790 [astro-ph.CO]}.

\bibitem{molinaparis}
C.~Molina-Par\'is and M.~Visser, ``Minimal conditions for the creation of a Friedman-Robertson-Walker universe from a `bounce'," Phys. Lett. B {\bf 455}, 90 (1999); {\tt gr-qc/9810023}.

\bibitem{Penrose:1965}
R.~Penrose, ``Gravitational Collapse and Space-Time Singularities," Phys. Rev. Lett. {\bf 14}, 57 (1965).
  
\bibitem{Farhi:1987}
E.~Farhi and A.~H.~Guth, ``An Obstacle to Creating a Universe in the Laboratory," Phys. Lett. B {\bf 183}, 149 (1987).
  
\bibitem{Hawking:1992} 
  S.~W.~Hawking,
  ``The Chronology Protection Conjecture,''
  Phys. Rev. D {\bf 46}, 603 (1992).
  
\bibitem{Hawking:1976} 
  S.~W.~Hawking,
  ``Black Holes and Thermodynamics,''
  Phys. Rev. D {\bf 13}, 191 (1976).
  
\bibitem{Barcelo:2002}
  C.~Barcelo and M.~Visser,
  ``Twilight for the Energy Conditions?,''
  Int.\ J.\ Mod.\ Phys.\ D {\bf 11}, 1553 (2002); {\tt gr-qc/0205066}.
  
 \bibitem{Rubakov:2014} 
  V.~A.~Rubakov,
  ``The Null Energy Condition and its Violation,''
  Phys. Usp. {\bf 57}, 128 (2014); {\tt arXiv:1401.4024 [hep-th]}.
  
\bibitem{quintombounce}
Y.-F.~Cai, T.~Qiu, Y.-S.~Piao, M.~Li, and X.~Zhang, ``Bouncing Universe with Quintom Matter," JHEP {\bf 0710}, 071 (2007); {\tt arXiv:0704.1090 [gr-qc]}.

\bibitem{galileonbounce}
T.~Qiu, J.~Evslin, Y.-F.~Cai, M.~Li, and X.~Zhang, ``Bouncing Galileon Cosmologies," JCAP {\bf 1110}, 036 (2011); {\tt arXiv:1108.0593 [hep-th]}.

\bibitem{Buchbinder:2007ad} 
  E.~I.~Buchbinder, J.~Khoury, and B.~A.~Ovrut,
  ``New Ekpyrotic Cosmology,''
  Phys. Rev. D {\bf 76}, 123503 (2007); {\tt hep-th/0702154}.
  
  \bibitem{ghostbounce}
C.~Lin, R.~H.~Brandenberger, and L.~P.~Levasseur, ``A Matter Bounce by Means of Ghost Condensation," {\tt arXiv:1007.2654 [hep-th]}.
   
\bibitem{superbounce}   
M.~Koehn, J.~L.~Lehners, and B.~A.~Ovrut, ``A Cosmological Super-Bounce," Phys. Rev. D {\bf 90}, 025005 (2014); {\tt arXiv:1310.7577 [hep-th]}.
  
 \bibitem{jones} 
  S.~Chatterjee, D.~A.~Easson, and M.~Parikh,
  ``Energy Conditions in the Jordan Frame,''
  Class. Quant. Grav. {\bf 30}, 235031 (2013); {\tt arXiv:1212.6430 [gr-qc]}.
    
\bibitem{NECderivation}
M.~Parikh and J.-P.~van der Schaar, ``Derivation of the Null Energy Condition," {\tt arXiv:1406.5163 [hep-th]}.
  
\bibitem{CallanThorlacius}
C.~G.~Callan, Jr. and L.~Thorlacius, ``Sigma Models and String Theory," in {\em Particles, Strings and Supernovae: Proceedings of the Theoretical Advanced Study Institute in Elementary Particle Physics (TASI 88),} edited by A.~Jevicki and C.~I.~Tan (World Scientific, Teaneck, N. J., 1989), Vol. 2, 795.

 \bibitem{GSW1}
  M.~B.~Green, J.~H.~Schwarz, and E.~Witten,
 {\em Superstring Theory: Introduction} (Cambridge, London, 1987).
  
\bibitem{prebigbang}
M.~Gasperini and G.~Veneziano, ``The Pre-Big Bang Scenario in String Cosmology," Phys. Rept. {\bf 373}, 1 (2003); {\tt hep-th/0207130}.
 
\bibitem{G-bounce}
D.~A.~Easson, I.~Sawicki, and A.~Vikman, ``G-Bounce," JCAP {\bf 1111}, 021 (2011); {\tt arXiv:1109.1047 [hep-th]}.
  
 \bibitem{langlois}
D.~Langlois and A.~Naruko,``Bouncing Cosmologies in Massive Gravity on de Sitter,''
  Class. Quant. Grav. {\bf 30}, 205012 (2013); {\tt arXiv:1305.6346 [hep-th]}.
  
\bibitem{torsionbounce} 
  Y.~F.~Cai, S.~H.~Chen, J.~B.~Dent, S.~Dutta, and E.~N.~Saridakis,
  ``Matter Bounce Cosmology with the f(T) Gravity,''
  Class. Quant. Grav. {\bf 28}, 215011 (2011); {\tt arXiv:1104.4349 [astro-ph.CO]}.
  
\bibitem{stringinspiredbounces}
T.~Biswas, A.~Mazumdar, and W.~Siegel, ``Bouncing Universes in String-Inspired Gravity," JCAP {\bf 0603}, 009 (2006); {\tt hep-th/0508194}.

 \bibitem{Brandenberger:1993ef} 
  R.~H.~Brandenberger, V.~F.~Mukhanov, and A.~Sornborger, ``A Cosmological Theory without Singularities,'' Phys. Rev. D {\bf 48}, 1629 (1993); {\tt gr-qc/9303001}.
  
\bibitem{savmartinec}
S.~R.~Green, E.~J.~Martinec, C.~Quigley, and S.~Sethi, ``Constraints on String Cosmology,''
  Class. Quant. Grav.  {\bf 29}, 075006 (2012); {\tt arXiv:1110.0545 [hep-th]}.
  
\bibitem{miguel1}
L.~Cornalba, M.~S.~Costa, and C.~Kounnas, ``A Resolution of the Cosmological Singularity with Orientifolds,'' Nucl. Phys. B {\bf 637}, 378 (2002); {\tt hep-th/0204261}.
  
\bibitem{miguel2}
L.~Cornalba and M.~S.~Costa,``Time-dependent orbifolds and string cosmology,''
  Fortsch. Phys. {\bf 52}, 145 (2004); {\tt hep-th/0310099}.

\end{thebibliography}
\end{document}